  \documentclass[twocolumn,prl,aps]{revtex4}

 \usepackage[dvips]{graphicx}
 \usepackage[dvips]{graphics}

\newlength{\bxwidth}\bxwidth=0.8\textwidth

\begin{document}
\title{Resonant magnetic excitations at high energy in superconducting $\bf YBa_2Cu_3O_{6.85}$}
\author{S. ~Pailh\`es$^1$, Y. ~Sidis$^1$, P.~Bourges$^{1\ast}$, V.
Hinkov$^2$, A. ~Ivanov$^3$, C. Ulrich$^2$, L.P.~Regnault$^4$ and B. Keimer$^{2}$}

\affiliation{
$^1$ Laboratoire L\'eon Brillouin, CEA-CNRS, CE-Saclay, 91191 Gif sur Yvette, France.\\
$^2$ Max-Planck-Institut f\"ur Festk\"orperforschung, 70569 Stuttgart, Germany\\
$^3$ Institut Laue Langevin, 156X, 38042 Grenoble cedex 9, France.\\
$^4$ CEA Grenoble, DRFMC, 38054 Grenoble cedex 9, France.
}

\pacs{PACS numbers: 74.25.Ha  74.72.Bk, 25.40.Fq }

\begin{abstract}

A detailed inelastic neutron scattering study of the high
temperature superconductor $\rm YBa_2Cu_3O_{6.85}$ provides
evidence of new resonant magnetic features, in addition to the well
known resonant mode at 41 meV: (i) a commensurate magnetic
resonance peak at 53 meV with an even symmetry under exchange of
two adjacent $\rm CuO_2$ layers; and (ii) high energy
incommensurate resonant spin excitations whose spectral weight is
around 54 meV. The locus and the spectral weight of
these modes can be understood by considering the momentum shape of the 
electron-hole spin-flip continuum of $d$-wave superconductors. This provides new insight 
into the interplay between collective spin excitations and the continuum of electron-hole excitations.
\end{abstract}

\maketitle

The spin excitation spectrum of the copper oxide superconductors
is modified markedly upon entering the superconducting state. In
materials with transition temperatures, $\rm T_c$, exceeding about
50 K, a ``resonant mode'' emerges below $\rm T_c$ around the
wavevector ${\bf Q_{AF}}\equiv \rm (\pi/a,\pi/a)$ characteristic of antiferromagnetism (AF) in
the undoped parent compounds\cite{Sidis03}. Materials in which the mode has been
observed include cuprates with one $\rm Tl_2Ba_2CuO_{6+\delta}$
(Tl2201) \cite{he02}) and two ($\rm YBa_2 Cu_3 O_{6+x}$ \cite{rossat,fong00,dai01}, $\rm
Bi_2Sr_2CaCu_2O_{8+\delta}$ \cite{fong99}) $\rm CuO_2$ planes per
unit cell. Very recently, an analogous feature has been observed
in the single-layer material $\rm La_{2-x}Ba_x Cu O_{4}$ as well,
but details of its temperature dependence are not yet known
\cite{jmt04}. The results of angle resolved photoemission
spectroscopy (ARPES) \cite{arpes}, optical conductivity \cite{optcond}, and
tunneling \cite{zasadzinski01} experiments on $\rm
Bi_2Sr_2CaCu_2O_{8+\delta}$ have been interpreted as evidence of
strong interactions of charged quasiparticles with this magnetic
mode. The mode therefore plays a prominent role in current
theories of high temperature
superconductivity \cite{abanov,norman}. Most models proposed to
explain the resonant mode are based on strong electronic
correlations
\cite{vdM,Liu95,Millis96,abanov,norman,Norman-FS,Onufrieva02,eremin,morr,demler,Vojta,Morais-smith,Batista,Kruger03}.
In an itinerant-magnetism picture, the mode is assigned to an
excitonic bound state in the superconductivity-induced gap in the
spectrum of electron-hole spin-flip excitations
\cite{vdM,Liu95,Millis96,abanov,norman,Norman-FS,Onufrieva02,eremin}. In
local-moment models, the mode is viewed as a magnon-like
excitation characteristic of a magnetically ordered phase
competing with the superconducting state
\cite{morr,demler,Vojta,Morais-smith,Batista,Kruger03}.

Early experiments on the resonant \cite{rossat} mode were focused on the
wavevector ${\bf Q_{AF}}$, where its spectral weight is maximum. Recent
advances in neutron scattering instrumentation have made it
possible to resolve weaker features of the spin excitations, which
further constrain the theoretical description of the mode and its
interaction with charged quasiparticles. First, incommensurate
excitation branches both below and above 41 meV\cite{Bourges-science,Reznik03}
were found to merge at ${\bf Q_{AF}}$, forming continuous dispersion 
relations with an X-shape. In
addition a second, disconnected mode was recently observed in
slightly overdoped $\rm Y_{0.9}Ca_{0.1}Ba_2Cu_3O_7$ ($\rm T_c$=85
K) \cite{Pailhes03}. Two magnetic modes with odd and even symmetry
with respect to exchange of two adjacent $\rm CuO_2$ layers are
expected on general grounds in a bilayer material such as $\rm
(Y,Ca)Ba_2Cu_3O_7$. In the overdoped material, the odd and even
modes were observed at 36 and 43 meV, respectively. The energy
difference is in good agreement with the interlayer exchange
interaction, $J_{\perp} \sim 10$ meV, extracted from the spin wave
dispersions in antiferromagnetically ordered $\rm
YBa_2Cu_3O_{6.15}$ \cite{Reznik96}.

In slightly underdoped $\rm YBa_2Cu_3O_{6.85}$, only the odd mode has
thus far been observed at an energy of 41 meV. A survey of the
high energy spectrum has now revealed a weak, even counterpart at
53 meV. While the even mode was expected to be present on general
grounds, the survey has also uncovered an unexpected incommensurate
excitation branch separated from both of these modes by a
``silent band'', where spin excitations are not observed. The
presence of silent bands in the spectrum of collective spin
excitations can be understood by considering the momentum shape of
the Stoner continuum of spin-flip excitations in a $d$-wave
superconductor.

The experiment was performed on the same sample used in Ref.
\cite{Bourges-science}, and the measurements were taken at the
recently upgraded IN8 triple axis spectrometer at the Institute
Laue Langevin (Grenoble, France). The IN8 beam optics include 
vertically and horizontally focusing monochromator
and analyzer, corresponding of arrays of Cu (200) crystals and  pyrolytic graphite (002)  
crystals, respectively. The measurements were performed
with a fixed final neutron energy of 35 meV. A filter was inserted
into the scattered beam in order to eliminate higher order
contamination. The crystal was oriented such that momentum
transfers $\bf Q$ of the form ${\bf Q}\rm =(H,H,L)$ were accessible. We use
a notation in which $\bf Q$ is indexed in units of the tetragonal
reciprocal lattice vectors $2\pi/a=1.64 $\AA$^{-1}$ and
$2\pi/c=0.54 $\AA$^{-1}$. Excitations of even (e) and odd (o)
symmetry can be identified by virtue of their magnetic structure
factors along the $c$-direction. As discussed in
Ref.~\cite{Pailhes03}, the neutron scattering cross section can be
written as,

\begin{eqnarray}
\frac{ \partial^2 \sigma ({\bf Q},\omega)}{\partial \Omega \partial
\omega} \propto f^2(Q) \Big[ \sin^2 (\pi z L)
 Im \lbrack \chi_{o}({\bf Q},\omega)\rbrack + \nonumber\\
 \cos^2(\pi z L) Im \lbrack
\chi_{e}({\bf Q},\omega)\rbrack \Big], \label{eq-bilayer}
\end{eqnarray}


\noindent
where $zc$=3.3 \AA\ is the distance between $\rm CuO_2$
planes within a bilayer unit, and $\rm f(Q)$ is the magnetic form
factor of $\rm Cu^{2+}$.

\begin{figure}[t]
\includegraphics[width=6.7cm,height=7cm]{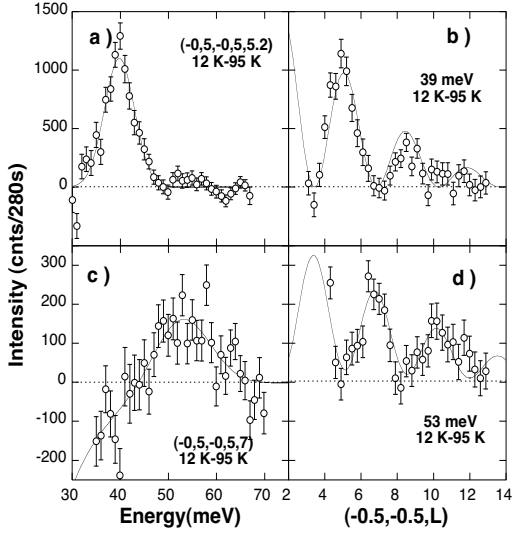}  
\caption {
{\label{fig1}} Differences between energy scans
measured at 12 K ($\rm <T_{c}$) and 95 K ($\rm >T_{c}$):
a) in the odd channel at $(-0.5,-0.5,5.2)$, c) in the even channel at $(-0.5,-0.5,7)$.
Differences between  10K and 95K of constant energy scan along the (001) direction:
b) at 39 meV, d) at 53 meV. The lines in b) follow the sine-square modulation  and in
d) the cosine-square modulation of Eq. \ref{eq-bilayer} convolved by the spectrometer 
resolution function. }
\end{figure}

\begin{figure}[t]
\centerline{\includegraphics[width=7 cm]{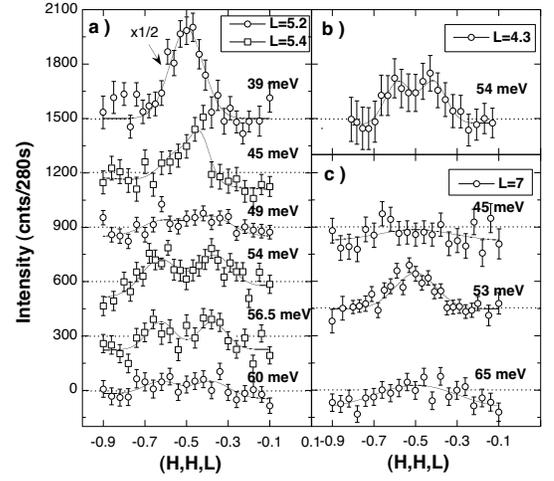}}
\caption{
Differences between 12 K ($\rm <T_{c}$) and 95 K ($\rm >T_{c}$)
of constant energy scans measured along the (110) direction
around (-0.5,-0.5,L):
a) L=5.2 or 5.4 (odd channel), b) L=7 (even channel), c) L=4.3
(mixed channel). For the sake of clarity, scans are shifted by 300 counts
with decreasing energy from bottom to top.
}
\label{fig2}
\end{figure}

Fig.~\ref{fig1}a shows the difference between constant-$\bf Q$ scans
measured at 12 K ($\rm <T_{c}$) and 95 K ($\rm >T_{c}$) and at the
wave vector (-0.5,-0.5,5.2) \cite{normalstate}. The enhancement of
the scattering intensity at $\rm E_r^o$=41 meV below $\rm T_c$
heralds the well know resonance peak in the odd channel.
Constant-energy scans at 39 meV indicate that the resonant
excitation is centered at $\bf Q_{AF}$ (Fig.~\ref{fig2}a) and
display the expected [$\rm f^2(Q) \sin^2 (\pi z L)$] modulation
along c* (Fig.~\ref{fig1}b). 
Since the dispersion of the odd mode has been well documented, it was not studied in
detail, but constant-energy and constant-$\bf Q$ profiles with
excitation energies around $\rm E_r^o$ (Figs. 1 and 2) are in good
agreement with the dispersion relations \cite{dispersion}
described in Refs. \cite{Bourges-science,Reznik03}.
In particular, the asymmetric lineshape of the constant-energy cut
at 45 meV (Fig.~\ref{fig1}a) can be reproduced by assuming an
upward dispersion  similar to that reported for $\rm
YBa_2Cu_3O_{6.95}$ \cite{Reznik03}. The asymmetry is due to a
standard resolution focusing effect.

\begin{figure}[t]
\includegraphics[width=4cm,height=4cm]{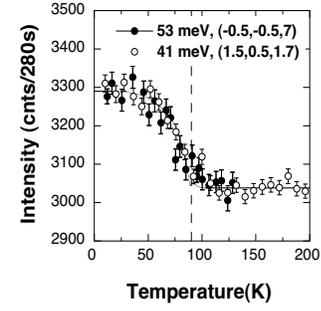}
\caption{
(full circles) Temperature dependence of the neutron
intensity at ${\bf Q_{AF}}=(0.5,0.5,7)$ at the even mode energy $E_r^e=53$ meV.
The intensity has been scaled at 12 K and 125 K to that of the odd mode at 41 meV (open circles), 
measured under different experimental conditions \cite{Bourges-science}.}
\label{fig3}
\end{figure}

In addition to the odd resonance, we discovered a second 
resonant magnetic excitation at higher energy. At $\bf Q_{AF}$=(-0.5,-0.5,7), 
where the structure factor for even
excitations is maximum, an enhancement of the response in the
superconducting state occurs at $\rm E_r^e$=53 meV
(Fig.~\ref{fig1}c). At this energy, the signal is peaked at the
AF wave vector (Fig.~\ref{fig2}c). Since its amplitude decreases with
increasing $\bf Q$ following the magnetic form factor, it can 
be identified as magnetic (Fig.~\ref{fig1}d). Furthermore, its
[$\rm f^2(Q) \cos^2 (\pi z L)$] modulation along c* demonstrates
that it has an even symmetry (Fig.~\ref{fig1}d). Finally, the
signal decreases precipitously at $\rm T_c$, as observed for the
magnetic resonance in the odd channel (Fig.~\ref{fig3}).

Even and odd resonance peaks also exhibit significant differences.
First, the enhancement of the magnetic response in the
superconducting state is six times weaker at $\rm E_r^e$ than at
$\rm E_r^o$. This explains why the even mode could not be
observed in prior experiments with weaker neutron flux. Second,
both the constant-energy and the constant-$\bf Q$ profiles of the even
spin excitation are broader than those of the odd mode. Gaussian fits of
constant-energy scans provide an intrinsic
full-width-at-half-maximum (FWHM) of  $\rm \Delta Q
=0.41\pm0.05$\AA$^{-1}$ at 53 meV in the even channel
(Fig.~\ref{fig2}c) whereas $\rm \Delta Q =0.25\pm0.05 $\AA$^{-1}$
at 39 meV in the odd channel (Fig.~\ref{fig2}a). In energy, the
even resonance peak displays an intrinsic FWHM of 11 meV, in contrast to
the resolution-limited odd mode. We performed a series of
constant-energy scans in the even channel at energies above and
below $\rm E_r^e$, but could not resolve a dispersion analogous to
that of the odd mode. We note, however, that the energy and wave vector
resolutions at 53 meV are significantly broader than at 41 meV.
Further study is required to ascertain whether the comparatively large energy and 
$Q$-widths of the even profiles are due to an intrinsic damping of the
mode, or due to an unresolved dispersion relation. Leaving these details aside, it is 
interesting to note that the energies of even and odd resonance peaks deviate by $\pm$6 meV
from 47 meV, the resonance peak energy of the mono-layer system
$\rm Tl_2Ba_2CuO_{6+\delta} $\cite{he02} at optimal doping. This splitting can be 
understood as a consequence  of the interlayer AF exchange coupling within a bilayer unit,
which is known to be of similar magnitude ($J_\perp \sim $10 meV).

\begin{figure}[b]
\centerline{\includegraphics[width=8.5 cm]{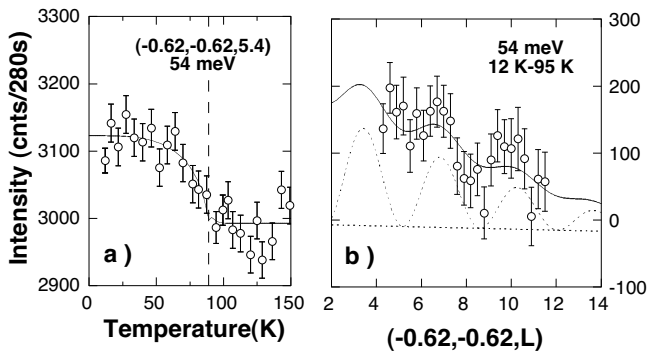}} \caption{
Temperature dependence of the neutron intensity at the odd wave
vector (-0.62,-0.62,5.4) and at $E=54$ meV. b) Difference between
constant energy scans at 12 K and 95 K along the (001) direction.
The full line is a fit of the data by a combination of the odd and
even modulations (see text). For clarity, the contribution from
the commensurate even resonant mode is represented by the dashed
line. } \label{fig4}
\end{figure}

Surprisingly, the survey of the magnetic spectrum in the energy
range 50-60 meV has uncovered another set of excitations
(Fig.~\ref{fig2}a) distinct from the commensurate even mode. Constant-energy 
scans show clear symmetric incommensurate magnetic excitations at 
$\bf Q_{IC}$=(-0.5$\pm \delta$, -0.5$\pm \delta$,L) with $\delta=$0.12 r.l.u both in 
the odd channel at $L=5.4$ (Fig.~\ref{fig2}a) and in the mixed channel at $L=4.3$(Fig.~\ref{fig2}a).
The temperature dependence of the IC peak intensity at $\bf Q_{IC}$=(-0.62,-0.62,5.4) and 54 meV 
displays a marked decrease when the temperature is raised above $\rm T_c$ (Fig.~\ref{fig4}.a), as 
previously reported for both odd and even resonance peaks (Fig.~\ref{fig3}). In addition, the IC peak 
intensity decreases along c* following $\rm f^2(Q)$, as expected for a magnetic signal Fig.~\ref{fig4}b). 
It is not straightforward to deduce the symmetry of the IC resonant excitations from the $L$-dependence 
measured at $\bf Q_{IC}$, because it overlaps with the cosine-square modulation of the nearby even 
resonance peak. To overcome this difficulty, we fitted all of the data around 53-54 meV
reported in  Figs.~\ref{fig1}d,~\ref{fig2}a-c, and ~\ref{fig4}b simultaneously, assuming an 
L-dependence for the IC spin excitation of the form 
[$\rm f^2(Q)(\gamma \sin^2(\pi zl) + (1-\gamma) \cos^2(\pi zl))$].
The best fit is given by $\rm \gamma = 0.8 \pm 0.15$, which means that
these incommensurate excitations 
exhibit predominantly an odd symmetry. The intensity of the IC spin fluctuations passes 
through a maximum around 54 meV, where it reaches about 1/7 of the odd resonance intensity 
(Fig.~\ref{fig2}a). In the energy range from $\sim$ 50 meV to $\sim$ 60 meV,
the position of the IC peaks, $\rm \delta \simeq$ 0.12 r.l.u., is energy independent 
within the error and the IC peaks are symmetric on both sides of $\bf Q_{AF}$, 
in contrast to the resolution focusing effect observed at 45 meV (Fig.~\ref{fig2}a).
Interestingly, the IC intensity drops around 49 meV separating both behaviors (Fig.~\ref{fig2}a).

\begin{figure}[t]
\includegraphics[width=7cm,height=5cm]{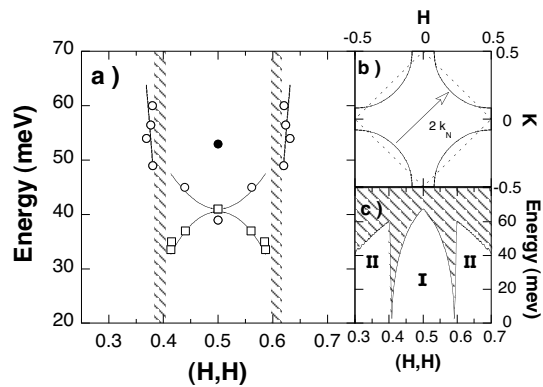}
\caption{
a) Energy distribution of the resonant spin excitations along the (110) direction:
open (full) circles stand for excitation with odd (even) symmetry (from Fig.~\ref{fig2}).
The downward dispersion observed in \cite{Bourges-science} is reproduced, assuming it
is isotropic (open squares). The vertical hatched areas indicate the locus of the
``silent band'' described in the text. b) Typical Fermi surface measured
by ARPES in BSCO \cite{Kordyuk03,Norman-FS}. c) Electron-hole spin-flip continuum along the (110)
direction of a $d$-wave superconductor with a maximum superconducting gap of 35 meV. The continuum goes
to zero energy for 2$\bf k_N$= (0.5$\pm\delta$,0.5$\pm\delta$) with
$\delta \simeq$0.11 r.l.u. joining the nodal points at the Fermi surface (arrow in b).}
\label{fig5}
\end{figure}

Fig. 5a provides a synopsis of all spin excitation branches revealed in this study. The figure suggests 
that the high energy IC excitation branch is a continuation of the upper dispersion branch emanating 
from the odd resonance at 41 meV. However, the data of Fig. 2 demonstrate that the intensity along 
this branch is dramatically reduced around 49 meV and $\delta \sim 0.1-0.11$. This ``dark region'' is 
marked by the hatched area in Fig. 5a. It had been noticed before that the lower excitation branch 
emanating from 41 meV also fades out rapidly as it approaches $\delta \sim 0.1$ \cite{Bourges-science}. 
Interestingly, therefore, the spin fluctuation intensity is very weak in a narrow ``silent band'' around 
$\delta \sim 0.1-0.11$ over the entire energy range (Fig.\ref{fig5}a). 

We will argue now that the presence of ``silent bands'', the upper cutoff of the IC spin excitations, and 
the intensity ratio of odd and even modes can be understood by considering the shape of the electron-hole 
spin flip continuum of a $d$-wave superconductor. To illustrate this point, Fig.\ref{fig5}c displays the 
SC $d$-wave continuum along the (110) direction for the Fermi surface shown in Fig.\ref{fig5}b. Realistic 
calculations show that collective modes within the continuum or the continuum itself is likely much too 
weak to be directly observed by inelastic neutron scattering experiments. Therefore, only collective modes 
located below the {\it e-h} continuum can actually be detected. The continuum threshold exhibits a steep 
minimum in the vicinity of the wave vector 2$\bf k_N$ corresponding to scattering between nodes of the 
$d$-wave gap function (Fig.\ref{fig5}b). This provides a natural explanation for the ``silent bands'' 
of Fig.\ref{fig5}a. 

In the same picture, the high energy IC excitations can be understood as collective modes in zone II of 
the gap in the electron-hole spin-flip continuum, whereas the odd and even excitation branches emanating 
from $\bf Q_{AF}$ are collective states in zone I. The upper cutoff of the IC excitation branch can then 
be identified with the energy threshold, $\hbar \omega_c$, of the continuum in zone II. An upper 
cutoff of 60 meV is well reproduced by assuming a maximum superconducting gap of 35 meV (Fig.\ref{fig5}c), 
consistent with the gap values extracted from ARPES and Raman scattering measurements around optimum doping.
As discussed in Refs. \onlinecite{Pailhes03,Millis96}, an independent estimate of the continuum threshold 
in zone I can be obtained through the intensity ratio of the odd and even modes at $\bf Q_{AF}$. In theories 
according to which the resonant spin excitations correspond to collective S=1 modes pulled below the 
continuum by antiferromagnetic interactions, the spectral weights of the resonant modes are approximately 
proportional to their binding energies, $\hbar \omega_c - E_r^e$ and $\hbar \omega_c - E_r^o$. 
Depending whether energy-integrated intensities or peak intensities are compared, the spectral weight 
ratio is either 3 or 6, yielding $\hbar \omega_c=$ 66 meV and 55 meV, respectively. Fig. \ref{fig5}c shows 
that the estimate  $\hbar \omega_c$ obtained in this way is in good agreement with that from the 
high-energy IC excitations discussed above.

In conclusion, our inelastic neutron scattering experiment in $\rm YBa_2Cu_3O_{6.85}$ has uncovered a
commensurate magnetic resonance peak of even symmetry at 53 meV; its intensity is 6 times weaker
than the well-known odd mode at 41 meV. Further, we have found high-energy spin excitations with a 
weakly energy-dependent incommensurability, separated by a ``silent band'' from the 41 meV mode. The 
silent band and other features of the spin excitation spectrum can be understood as fingerprints of 
the interaction of collective spin excitations with electron-hole excitations. They can thus be regarded 
as counterparts of the self-energy effects of charged quasiparticles revealed in various spectroscopies
\cite{arpes,optcond,zasadzinski01,abanov,norman}. A quantitative description of the excitation spectrum 
of Fig. \ref{fig5}a is a challenge for both itinerant and local-moment theories of magnetic excitations 
in the cuprates.


\begin{thebibliography}{10}
\bibitem{normalstate} As previously reported \cite{Bourges-science}, magnetic
excitations,  broad in both momentum and energy, are observed in the normal state in
that sample. Following the standard technique, we perform a difference of the neutron
intensity between $\sim$ 12 K and  $\sim$ 100 K in order to identify the resonant enhancements
of the spin susceptibility.
\bibitem{dispersion}  The downward (-) and upward (+) dispersions  can be  modelized by the
relation: $\rm E_{\pm}=\sqrt{(E^o_r)^2\pm(\alpha_\pm q)^{2})}$ with
$\rm E_r$=40.5 meV,
$\rm \alpha _{-} \simeq$ 125 meV.\AA\ and
$\rm \alpha _{+} \simeq$ 110 meV.\AA.
\end{thebibliography}
\end{document}